\documentclass[sigconf]{acmart}

\usepackage[utf8]{inputenc} 
\usepackage[T1]{fontenc}    
\usepackage{enumitem}
\usepackage{url}
\usepackage{siunitx}
\usepackage{multirow}
\usepackage{xspace}
\usepackage[capitalise]{cleveref}
\usepackage{subcaption}
\usepackage{listings}
\usepackage{soul}
\usepackage{amsfonts}
\usepackage{nicefrac}
\usepackage{microtype}
\usepackage{arydshln}
\usepackage{textcomp}

\definecolor{codegray}{rgb}{0.5,0.5,0.5}
\definecolor{pblue}{rgb}{0.13,0.13,1}
\definecolor{pgreen}{rgb}{0,0.5,0}
\definecolor{pred}{rgb}{0.9,0,0}
\definecolor{pgrey}{rgb}{0.46,0.45,0.48}
\definecolor{back_color}{rgb}{0.95,0.95,0.92}

\AtBeginDocument{%
  }

\copyrightyear{2026}
\acmYear{2026}
\setcopyright{cc}
\setcctype{by}
\acmConference[FORGE '26]{2026 IEEE/ACM Third International Conference on AI Foundation Models and Software Engineering}{April 12--13, 2026}{Rio de Janeiro, Brazil}
\acmBooktitle{2026 IEEE/ACM Third International Conference on AI Foundation Models and Software Engineering (FORGE '26), April 12--13, 2026, Rio de Janeiro, Brazil}
\acmPrice{}
\acmDOI{10.1145/3793655.3793724}
\acmISBN{979-8-4007-2477-0/2026/04}

\begin{document}

\title{Deep Graph-Language Fusion for Structure-Aware Code Generation}

\author{Mert Tiftikci}
\authornotemark[1] \authornotemark[2]
\email{mert.tiftikci@tu-darmstadt.de}
\affiliation{%
  \institution{Technische Universit\"{a}t Darmstadt}
  \city{Darmstadt}
  \country{Germany}
}

\author{Amir Molzam Sharifloo}
\email{amir.molzam@tu-darmstadt.de}
\affiliation{%
  \institution{Technische Universit\"{a}t Darmstadt}
  \city{Darmstadt}
  \country{Germany}
}

\author{Mira Mezini}
\email{mezini@informatik.tu-darmstadt.de}
\affiliation{%
  \institution{Technische Universit\"{a}t Darmstadt}
  \city{Darmstadt}
  \country{Germany}
}
\additionalaffiliation{%
  \institution{The Hessian Center for Artificial Intelligence (hessian.AI)}
  \country{Germany}
}
\additionalaffiliation{%
  \institution{National Research Center for Applied Cybersecurity ATHENE}
  \country{Germany}
}

\begin{abstract}

Pre-trained Language Models (PLMs) have the potential to transform software development tasks.
However, despite significant advances, current PLMs struggle to capture the structured and relational attributes of code, such as control flow and data dependencies.
This limitation is rooted in an architectural mismatch: whereas code structure is best represented by graphs, transformer-based LLMs process input as sequential token patterns and therefore lack explicit structural awareness.
While recent research has explored integrating graph-based code representations using techniques like graph feature extraction, retrieval-augmented generation, and prompt engineering, existing approaches suffer from information loss during dense feature extraction or prompt encoding; notably, the potential of deep, token-level fusion of graph features within model internals has not been systematically explored.

In this paper, we initiate such an exploration by introducing CGFuse\footnote{The implementation is available at \url{https://github.com/stg-tud/cgfuse}.}, a novel framework that enables token-level integration of graph-derived representations by infusing learned graph features directly into the intermediate layers of pre-trained language models.
CGFuse combines a graph neural network (GNN) with a language model to explicitly preserve and exploit fine-grained structural information from code graphs, including abstract syntax trees and data-flow graphs.
We systematically evaluate CGFuse across multiple LLMs, demonstrating up to 10--16\% BLEU and 6--11\% CodeBLEU improvements in code generation performance.
These results highlight the potential of deep graph-PLM integration to advance the field toward more robust, capable AI-driven software development. 

\end{abstract}



\maketitle

\begin{figure*}[t]
    \centering
    \begin{subfigure}[t]{0.48\textwidth}
        \centering
        \begin{lstlisting}[language=Java, escapechar=!]
/**
 * Actually walks the bag to make sure the
 * count is correct and resets the running total
 * @return the current total size
 */\end{lstlisting}
        \caption{Example documentation.}
        \Description{Example documentation.}
        \label{fig:concode_input}
    \end{subfigure}
    \hfill
    \begin{subfigure}[t]{0.48\textwidth}
        \centering
        \begin{lstlisting}[language=Java, escapechar=!]
int calcTotalSize() {
    _total = extractList().size();
    return _total;
}\end{lstlisting}
        \caption{Example Java code.}
        \Description{Example Java code.}
        \label{fig:concode_output}
    \end{subfigure}
    \begin{subfigure}[t]{0.95\textwidth}
        \centering
        \includegraphics[width=\textwidth]{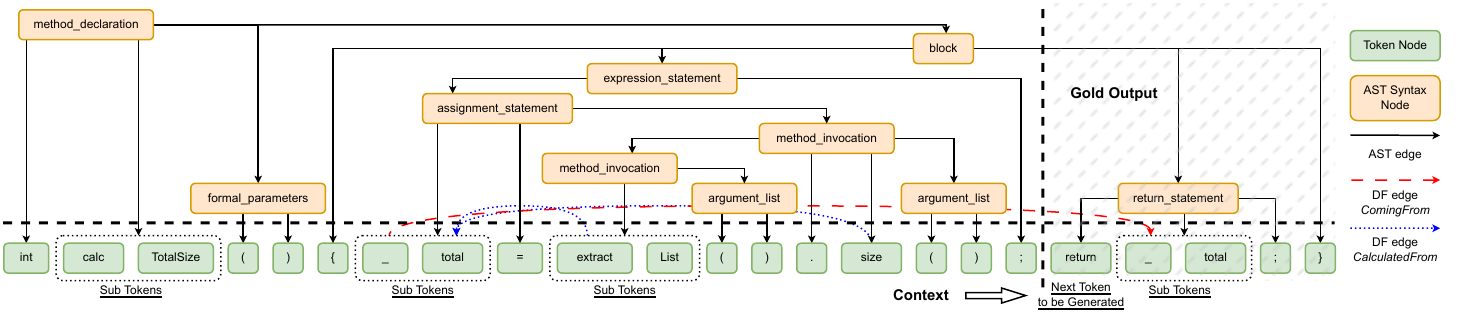}
        \caption{Code graph with AST (solid lines) and DFG edges (dashed/dotted). Syntax nodes $s$ are orange, terminal nodes $t$ are green.}
        \Description{Code graph with AST (solid lines) and DFG edges (dashed/dotted). Syntax nodes $s$ are orange, terminal nodes $t$ are green.}
        \label{fig:code_graph}
    \end{subfigure}
    \caption{A sample Java snippet with documentation (a), code (b), and its augmented code graph (c).}
    \Description{A sample Java snippet with documentation (a), code (b), and its augmented code graph (c).}
    \label{fig:input_example}
\end{figure*}

\section{Introduction}
\label{sec:introduction}

Large Language Models (LLMs) are rapidly transforming software engineering practice. State-of-the-art systems such as ChatGPT, Gemini, and CodeLLaMA now deliver significant capabilities in code generation, bug detection, and automated testing.
Their integration is advancing the field toward fully automated software development~\citep{maSWEGPTProcessCentricLanguage2025, xiaDemystifyingLLMBasedSoftware2025}.
However, recent studies reveal that, despite their strong performance on many tasks, these models struggle to explicitly capture the structured and relational nature of code, such as control flow and data dependencies~\citep{anandCriticalStudyWhat2024,WeiTOSEM24}.
We posit that this limitation stems from a fundamental architectural mismatch: while graph-based representations naturally encode structural relationships, transformer architectures primarily operate on sequential token patterns, providing only implicit structural awareness. 

To bridge this gap, several approaches have been proposed that integrate graph-based code representations through techniques such as graph-based feature extraction, retrieval-augmented generation (RAG), and prompt engineering. 
Some approaches operate at the repository level, creating complex hybrid code graphs that incorporate project file structures and sophisticated RAG systems~\citep{liuGraphCoderEnhancingRepositoryLevel2024, taoCodeGraphModel2025}.
At the token-and-model level, earlier work such as GraphCodeBERT~\citep{guoGraphCodeBERTPretrainingCode2020} and UniXcoder~\citep{guo-etal-2022-unixcoder} incorporate Abstract Syntax Trees (ASTs)\citep{knuth1968semantics} and data-flow graphs by flattening them as prompts and adding graph-based auxiliary training objectives for full pertaining.
More recent approaches \citep{zhangGALLaGraphAligned2024, duCodeGRAGBridgingGap2025} employ soft prompting, wherein GNN-generated features are appended to the input prompt and the model is fine-tuned accordingly.
While this approach enables direct learning from graphs, it incurs information loss during dense feature extraction and prompt encoding, limiting the depth of graph-LLM fusion.
Pass-Tuning~\citep{chenPassTuningStructureAwareParameterEfficient2023} advances this direction by using GNN-based experts within a prefix-tuning setup to achieve deeper fusion at the model level.
However, it still relies on dense graph representations, which constrains the model’s capacity to capture rich and fine-grained relational code patterns.

Notably, existing approaches have not systematically explored token-level fusion of extracted graph features.
This gap raises an important research question: can deeper integration achieved by infusing graph knowledge directly into the internal layers of pre-trained language models (PLMs) better preserve and exploit fine-grained structural information to enhance model capabilities on coding tasks?
Towards addressing this question, we developed \textbf{CGFuse}, a novel framework that seamlessly integrates graph-based representations of code into PLMs to enhance their understanding and generation capabilities.
Our approach trains a graph neural network to process code graphs and infuses its learned representations directly into the intermediate layers of PLMs, enabling token-level fusion. Specifically, our main contributions are as follows:

\noindent
\textbf{Novel Framework for Deep Graph-PLM Integration:} We propose \textbf{CGFuse}, the first framework that systematically fuses graph neural networks (GNNs) with pre-trained language models (PLMs) at intermediate layers, enabling PLMs to directly exploit fine-grained structural and relational information from code graphs such as ASTs and control/data-flow graphs at the token level.

\noindent
\textbf{Systematic Evaluation Across Models and Architectures:}
We conduct a systematic evaluation across nine selected PLMs -- including Encoder and Encoder-Decoder architectures -- and three GNN architectures, demonstrating that \textbf{CGFuse} improves code generation performance and effectively leverages structural information.

\section{Our Approach}
\label{sec:methodology}

We aim to enhance code generation quality by combining GNNs' structural understanding with transformer-based PLMs' contextual knowledge.
Our approach consists of three stages: (1) constructing code graphs from source code, (2) pre-training GNNs as graph experts, and (3) fusing GNN representations into a PLM for downstream tasks.

\subsection{Constructing Code Graphs}
\label{sect:code_graph}

To capture both syntactic and semantic dependencies in code, we represent each snippet as a \textit{code graph} $\mathcal{G}_c$. This graph is built by augmenting an Abstract Syntax Tree (AST) with data flow edges \citep{guoGraphCodeBERTPretrainingCode2020}. The AST
encodes the syntactic structure of the program, while data flow edges model how values propagate through variables.

Figure~\ref{fig:input_example} illustrates a sample Java snippet and its corresponding graph. Terminal nodes (\(\mathcal{T}\), green) represent code tokens, while syntax nodes (\(\mathcal{S}\), orange) capture higher-level constructs such as expressions and statements. Edges include AST hierarchy edges (solid lines) and data flow edges (dashed/dotted lines) labeled \textit{comingFrom} and \textit{calculatedBy}, representing the origin and computation of variables, respectively. Formally, the graph is defined as:
\[
\mathcal{G}_c = (\mathcal{V}, \mathcal{E}, \mathcal{R}), \quad \mathcal{V} = \mathcal{S} \cup \mathcal{T}, \quad \mathcal{R} = \{p, co, ca\},
\]
where $p$ denotes AST edges, $co$ denotes \textit{comingFrom}, and $ca$ denotes \textit{calculatedBy} edges.

\subsection{Graph Expert Pre-training}
\label{sec:gnn_pretraining}

We pre-train a GNN to serve as a graph expert, learning node-level representations for both syntax and terminal nodes. The GNN aggregates features from each node’s neighborhood $\mathcal{N}(v)$:
\begin{equation}
    \begin{split}
        a_v^l &= \texttt{aggregate}^l(\{h_u^{l-1} : u \in \mathcal{N}(v)\}), \\
        h_v^l &= \texttt{combine}^l(h_v^{l-1}, a_v^l),
    \end{split}
    \label{eq:gnn_aggregation_combine}
\end{equation}
where $h_v^0$ are initial node embeddings, which include PLM token embeddings for terminal nodes and learned embeddings for syntax nodes. For graphs with multiple relation types (AST, \textit{comingFrom}, \textit{calculatedBy}), we optionally use relational aggregation \citep{hamiltonInductiveRepresentationLearning2017}.

The GNN is trained on a node classification task, learning to predict properties of each node from its local context, capturing structural and semantic code patterns.
We initialize the GNN with PLM embeddings for terminal nodes and random embeddings for syntax nodes, and pre-train on node classification, masking a portion of nodes for validation/testing.

\subsection{Fusing GNN and PLM Features}
\label{sect:fusing}

\begin{figure}[t]
    \centering
    \includegraphics[width=\linewidth]{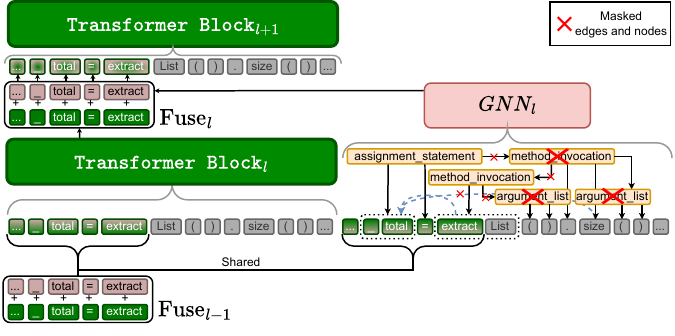}
    \caption{Fusion of GNN features into a PLM layer. Only context nodes are included and edges outside the context are masked.}
    \Description{Fusion of GNN features into a PLM layer. Only context nodes are included and edges outside the context are masked.}
    \label{fig:fusing}
\end{figure}

After pre-training, we integrate GNN embeddings into a transformer-based PLM. Figure~\ref{fig:fusing} illustrates this fusion mechanism. At a target layer $l$, the transformer produces token representations $h_\theta^l$, while the GNN outputs node representations $h_g^l$. We fuse them as:
\begin{gather}
     h_g^l = \texttt{GNN}^l(\mathcal{G}_c^l), \quad h_\theta^l = \theta^l(h_{f}^{l-1}), \quad h_f^l = h_\theta^l + \lambda h_g^l,
    \label{eq:fusing}
\end{gather}
where $\lambda$ controls fusion strength. Only nodes in the current context are fed into the GNN; out-of-context edges are masked. 

As for the implementation, we first attach the pre-trained GNN to target PLM layers and train it alone for several warmup epochs. Then, we jointly fine-tune the fused model on the downstream task, optimizing both PLM and GNN parameters.

\section{Experiments}
\label{sec:experiments}

In this section, we first present the experimental setup, including the datasets, model configurations, and evaluation metrics used to assess the effectiveness of our approach.
We then describe the experiments and report the results to demonstrate the impact of incorporating graph-based representations on model performance.

\subsection{Experimental Setup}
\label{sec:setup}

\noindent
\textbf{Baseline Models:} \label{sect:baselines}
We apply our approach to various architectures, trained with natural language, code, and flattened AST inputs.
\cref{tab:models} provides a comprehensive list of the models used in our experiments.
We construct encoder-decoder models for our experiments by adapting encoder-only models.
As a result, the parameter counts reported for these models correspond to their encoder-decoder versions, which approximately double the parameters of the original encoder models.

\noindent
\textbf{GNN experts:} \label{sect:gnn_models}
We investigated the impact of different GNN architectures and the number of intermediate layers.
Specifically, we experimented with the Relational Graph Convolutional Network (R-GCN)~\citep{hamiltonInductiveRepresentationLearning2017}, the GraphSAGE model (GS)~\citep{hamiltonInductiveRepresentationLearning2017}, and the Graph Isomorphism Network (GIN)~\citep{xuHowPowerfulAre2018}.

\noindent
\textbf{Dataset:} \label{sect:datasets}
For our experiments, we use the \textit{CONCODE} dataset\footnote{Dataset available at \url{https://huggingface.co/datasets/AhmedSSoliman/CodeXGLUE-CONCODE}.} -- an established benchmark for evaluating code generation models.
CONCODE comprises 104k Java classes (100k for training and 2k each for validation and testing) collected from GitHub. The code generation task takes a natural-language intent (see \cref{fig:input_example}) and generates code to fulfill it.

\noindent
\textbf{Metrics:} \label{sect:metrics}
For evaluation, we use three string-matching-based metrics:
\noindent\textit{Exact Match (EM)}: Measures the percentage of predictions that exactly match the ground truth.
\textit{BLEU (B)}: A widely used metric for evaluating text generation tasks, introduced by \citet{papineniBLEUMethodAutomatic2001}.
\textit{CodeBLEU (CB)}: An extension of BLEU tailored for code, which incorporates syntactic and semantic features~\citep{renCodeBLEUMethodAutomatic2020}.
It aggregates four complementary signals of n\mbox{-}gram match BLEU, weighted n\mbox{-}gram match that up-weights language keywords, syntax match computed over ASTs, and semantic match computed over DFGs, all contribute equally.

\noindent
\textbf{Implementation details:} \label{sect:implementaiton}
All experiments were conducted with a single run and the same seed value, using a single-layer GNN expert attached to the decoder's last layer.
All code graphs, except for sub-tokens, are generated before the training.
Sub-tokens are generated during model training and testing using the corresponding tokenizer.
The GNNs are pretrained for three epochs on these code graphs, which is sufficient to achieve approximately 99\% accuracy.

\begin{table}[t]
    \caption{Code Models evaluated in this work.
    We evaluate two types of architectures (Arch): encoder (\textbf{E}) and encoder-decoder (\textbf{ED}) models. In addition to natural language inputs (\textbf{NL}), models can be trained with code (\textbf{C}) or flattened AST inputs (\textbf{AST$_{f}$}), where the subscript indicates flattening. For all encoder models converted to encoder-decoder models, the number of parameters is specified for the encoder-decoder versions.
    }
    \label{tab:models}
    \centering
    \begin{tabular}{l c c c}
        \toprule
        Models & Param & Arch & Data \\
        \midrule
        BERT~\citep{devlinBERTPretrainingDeep2019} & 255M & E & NL \\
        RoBERTa~\citep{liuRoBERTaRobustlyOptimized2019b} & 285M & E & NL \\
        BART~\citep{lewisBARTDenoisingSequencetoSequence2019} & 143M & ED & NL \\
        T5~\citep{raffelExploringLimitsTransfer2020a} & 226M & ED & NL \\
        \midrule
        CodeBERT~\citep{feng2020codebert} & 277M & E & C \\
        GraphCodeBERT~\citep{guoGraphCodeBERTPretrainingCode2020} & 277M & E & C + DFG$_{f}$ \\
        UniXCoder~\citep{guo-etal-2022-unixcoder} & 280M & E & C + AST$_{f}$ \\
        PLBART~\citep{ahmadUnifiedPretrainingProgram2021a} & 139M & ED & C \\
        CodeT5~\citep{wangCodeT5IdentifierawareUnified2021a} & 222M & ED & C \\
        \bottomrule
    \end{tabular}
\end{table}

\subsection{Results}
\label{sec:results}

Our results in \cref{tab:concode_results} reveal several key observations:

\noindent
\textbf{Strong gains from GNN augmentation for natural language models:} Models pre-trained solely on natural language benefit substantially from GNN-based code graph infusion.
For example, BERT improves +15.1 BLEU (33.4 $\rightarrow$ 48.5) and +11.9 CodeBLEU (31.2 $\rightarrow$ 43.1) when combined with a 1-layer R-GCN.
This demonstrates that structural embeddings effectively infuse code knowledge into models that were not exposed to code during pre-training.

Selected NL models served as the base checkpoints from which the corresponding code-pretrained transformer models are initialized.
To assess sample efficiency, we apply \textbf{CGFuse} to RoBERTa and BART. The initialization chains are:
\begin{equation}
    \label{eq:initialization_chain}
    \begin{split}
        & \text{RoBERTa} \rightarrow \text{CodeBERT} \rightarrow \text{GraphCodeBERT}, \\
        & \text{RoBERTa} \rightarrow \text{UniXcoder}, \quad
        \text{BART} \rightarrow \text{PLBART}
    \end{split}
\end{equation}
Comparing the underlined entries in \cref{tab:concode_results} shows that \textbf{CGFuse}-enhanced NL models outperform their code-pretrained counterparts, despite using substantially fewer training samples and iterations (see \cref{tab:sample_efficiency})\footnote{A notable limitation of T5 \citep{raffel2020exploring} is its limited vocabulary. This makes CodeBLEU evaluation significantly unreliable; therefore, the T5 family is omitted from this comparison.}.

\begin{table}[t]
    \caption{Code generation results for baseline and fused models. For each fused model, the GNN architecture is specified with a subscript. $\lambda$ represents the fusing strength; see \cref{eq:fusing}.}
    \label{tab:concode_results}
    \centering
        \renewcommand{\arraystretch}{1.1}
        \begin{tabular}{c l l c c c}
            \toprule
            & Models & $\lambda$ & BLEU & CodeBLEU & EM \\
            \midrule
            \multirow{8}{*}{\rotatebox[origin=c]{90}{Natural Language}}
            & BERT & & 33.4 & 31.2 & 8.0  \\
            & BERT$_{R-GCN}$ & 1 & 48.5 & 43.1 & 10.1 \\
            \cdashline{2-6}
            & RoBERTa & & 36.8 & 33.1 & 11.6 \\
            & RoBERTa$_{GIN}$ & 0.5 & \underline{52.6} & \underline{45.0} & \underline{15.6} \\
            \cdashline{2-6}
            & BART & & 36.9 & 31.3 & 15.4 \\
            & BART$_{GIN}$ & 0.5 & \underline{52.1} & \underline{42.7} & \underline{20.4}  \\
            \cdashline{2-6}
            & T5 & & 33.5 & 29.9 & 15.2 \\
            & T5$_{GIN}$ & 0.5 & 40.0 & 31.7 & 19.5 \\
            \midrule
            \midrule
            \multirow{10}{*}{\rotatebox[origin=c]{90}{Programming Language}}
            & CodeBERT & & \underline{36.9} & \underline{34.4} & \underline{11.1} \\
            & CodeBERT$_{GIN}$ & 1 & 54.1 & 46.2 & 19.0 \\
            \cdashline{2-6}
            & GraphCodeBERT & & \underline{49.3} & \underline{35.4} & \underline{13.7} \\
            & GraphCodeBERT$_{GS}$ & 1 & 56.4 & 47.3 & 20.0 \\
            \cdashline{2-6}
            & UniXCoder & & \underline{40.5} & \underline{35.9} & \underline{14.0} \\
            & UniXCoder$_{GIN}$ & 1 & 57.5 & 49.5 & 21.2 \\
            \cdashline{2-6}
            & PLBART & & \underline{42.1} & \underline{36.6} & \underline{18.2} \\
            & PLBART$_{R-GCN}$ & 1 & 59.3 & 51.5 & 27.6 \\
            \cdashline{2-6}
            & CodeT5 & & 42.7 & 36.8 & 18.2 \\
            & CodeT5$_{GIN}$ & 1 & 60.0 & 51.5 & 30.3 \\
            \bottomrule
        \end{tabular}
\end{table}

\noindent
\textbf{Significant improvements for code-pretrained models:} PLBART, CodeBERT, GraphCodeBERT, and UniXCoder consistently improve across all metrics when augmented with a GNN.
In particular, PLBART shows greater improvement than the other models, with differences of +17.2, +14.9, and +9.4 on BLEU, CodeBLEU, and EM scores, respectively.
Overall, these results demonstrate that GNN-based code graph embeddings can improve code generation accuracy.

We further investigated how the number of layers in a GNN affects performance on the code-generation task.
We experimented with 1, 2, and 3 intermediate GNN layers to assess the effect of depth.
Each additional layer increases the node's receptive field, allowing aggregation from more distant neighbors, but it also adds parameters and increases training cost.
Results (\cref{table:gnn_results}) show that single-layer GNNs consistently outperform deeper variants across all tested architectures.
Adding a second or third layer leads to a substantial drop in scores, suggesting that deeper GNNs may introduce noise or over-smooth node representations, thereby reducing their usefulness for code generation.
Among the models, R-GCN achieves the highest scores in BLEU and CodeBLEU, whereas GIN achieves the highest score in EM, both with a single layer.
Overall, these results indicate that shallow GNN embeddings provide the most effective structural information for code generation.

\begin{table}[t]
    \centering
    \caption{The table shows the difference in training time (normalized \#Step), the amount of pretraining code data that models are exposed to (\#Sample), and the performance gains of NL-based models when CGFuse is applied, compared to their corresponding code model counterparts (e.g., GraphCodeBERT vs. RoBERTa$_{GIN}$; see \cref{eq:initialization_chain}).}
    \label{tab:sample_efficiency}
    \begin{tabular}{l l l l l l}
    	\toprule
        Models & $\times$\#Step & +\#Sample & -$\Delta$B & -$\Delta$CB & -$\Delta$EM \\
        \hline
		CodeBERT       & $\times$137 & $\sim$8M    & 15.7 & 10.6 & 4.5  \\
		GraphCodeBERT  & $\times$274 & $\sim$12M   & 13.3 &  9.6 & 1.9  \\
		UniXcoder      & $\times$547 & $\sim$36M   & 12.1 &  9.1 & 1.6  \\
		PLBART         & $\times$293 & $\sim$575M  & 10.0 &  6.1 & 2.2  \\
        \bottomrule
    \end{tabular}
\end{table}

\begin{table}[t]
    \caption{The average code generation results of CGFuse using different GNN architectures and numbers of layers on 9 PLMs at the last decoder layer ($\lambda = 1$). The numbers in the model names indicate the number of GNN layers (L).}
    \label{table:gnn_results}
    \centering
    \setlength{\tabcolsep}{1.1mm}{
    \begin{tabular}{l c c c | c c c | c c c}
        \toprule
        \multirow{2}{*}{Models} & \multicolumn{3}{c|}{BLEU} & \multicolumn{3}{c|}{CodeBLEU} & \multicolumn{3}{c}{EM} \\
        & 1-L & 2-L & 3-L & 1-L & 2-L & 3-L & 1-L & 2-L & 3-L \\
        \midrule
        R-GCN & \textbf{53.3} & 27.9 & 24.3 & \textbf{46.0} & 28.4 & 27.7 & 17.5 & 1.2 & 1.3 \\
        GS & 52.4 & 29.4 & 23.9 & 44.6 & 30.5 & 28.0 & 18.0 & 1.1 & 1.4 \\
        GIN & 52.0 & 14.1 & 12.1 & 44.3 & 20.4 & 20.0 & \textbf{20.0} & 1.7 & 2.0 \\
        \bottomrule
    \end{tabular}
    }
\end{table}

\section{Conclusion and Future Directions}
\label{sec:discussion}

In this work, we investigated how code PLMs can be enriched with representations generated by GNN experts at the token and layer levels. We evaluated our approach on various code and natural language PLMs in the code generation task, experimenting with nine different models pre-trained on both natural and programming languages, covering different architectures and training strategies. In general, we achieved better performance across all models. Moreover, the comparison between the NL models and their code counterparts shows significant sample efficiency in learning.

The encouraging results presented in the paper motivate further pursuing the research in several directions. 
First, developing strategies for efficient GNN training, handling incomplete code during decoding, and mitigating ambiguity in AST generation are promising avenues for further improvement.
Training a new GNN for each architectural or vocabulary change is computationally costly and integrating GNN experts into decoders for code generation tasks is challenging, as code graphs must often be constructed from incomplete or ambiguous code, complicating the generation process.
Second, applying the presented approach to larger PLMs could test its scalability and benefits for more complex generation tasks. Finally, evaluating performance on benchmarks that measure functional correctness such as HumanEval~\citep{chen2021evaluating} and MBPP~\citep{austinMBPP2021program} would provide a more rigorous assessment of practical utility.

\begin{acks}
We gratefully acknowledge support from the hessian.AI Service Center (funded by the Federal Ministry of Research, Technology and Space, BMFTR, grant no. 16IS22091) and the hessian.AI Innovation Lab (funded by the Hessian Ministry for Digital Strategy and Innovation, grant no. S-DIW04/0013/003) and also by the National Research Center for Applied Cybersecurity ATHENE within the project Athene SecureCoder, and by the LOEWE initiative (Hesse, Germany) [LOEWE/4a//519/05/00.002(0013)/95].
The work has benefited from the early stages of the funding by the Deutsche Forschungsgemeinschaft (DFG, German Research Foundation) under Germany’s Excellence Strategy— EXC-3057.
\end{acks}

\bibliographystyle{ACM-Reference-Format}
\bibliography{custom_citations}

\end{document}